\documentclass[twocolumn]{emulateapj}
\usepackage{graphicx}
\usepackage{epsfig}

\shorttitle{Neutron Star Atomic Features}
\shortauthors{Baub\"ock et al.}

\begin{document}
\title{Narrow Atomic Features from Rapidly Spinning Neutron Stars}
\author{Michi Baub\"ock, Dimitrios Psaltis, and Feryal \"Ozel}
\affil{Astronomy  Department,
University of Arizona,
933 North Cherry Avenue,
Tucson, AZ 85721, USA}

\email{email: mbaubock, dpsaltis, fozel@email.arizona.edu}

\begin{abstract}
Neutron stars spinning at moderate rates ($\sim$300-600~Hz) become oblate in shape and acquire a nonzero quadrupole moment. In this paper, we calculate profiles of atomic features from such neutron stars using a ray-tracing algorithm in the Hartle-Thorne approximation. We show that line profiles acquire cores that are much narrower than the widths expected from pure Doppler effects for a large range of observer inclinations. As a result, the effects of both the oblateness and the quadrupole moments of neutron stars need to be taken into account when aiming to measure neutron star radii from rotationally broadened lines.  Moreover, the presence of these narrow cores substantially increases the likelihood of detecting atomic lines from rapidly spinning neutron stars. 

\end{abstract}
\keywords{stars: neutron --- relativistic processes --- gravitation}

\section{Introduction}
Atomic features from neutron star surfaces, such as absorption lines, emission lines, and edges, provide some of the most direct probes of the physical properties of these objects. The origin of the features depends on the composition of the stellar atmosphere and the conditions at the stellar surface, such as the temperature, the metallicity, and the magnetic field strength.

Apart from probing the conditions of the stellar surface, it has long been recognized that atomic features can help constrain the neutron-star radius $R$ and its mass $M$. In particular, two global parameters of a neutron star can be measured from observations of such spectral features. The redshift of the line center can be used to determine the compactness $2GM/Rc^2$ of the neutron star. At the same time, the line width from a neutron star of known spin frequency can lead to a measurement of its radius (\"Ozel \& Psaltis 2003). Together, these two parameters can constrain the equation of state of neutron star interiors.

The detailed profile and overall redshift of atomic features from neutron-star surfaces is dictated primarily by the properties of their spacetimes. The spacetimes of slowly spinning neutron stars can be approximated by the Schwarzschild metric, with the addition of Doppler shifts due to the velocity of their surfaces (e.g., Miller \& Lamb 1998; Muno et al.\ 2002; Poutanen \& Beloborodov 2006). In this approximation, the rotational broadening of spectral features is largely due to the Doppler shift and is approximately proportional to the neutron star radius (\"Ozel \& Psaltis 2003; Chang et al. 2006). 

The Schwarzschild approximation neglects three effects which become important for more rapid rotation. To first order in stellar spin, the rotation of the star causes frame dragging in the spacetime around it. Bhattacharyya et al.\ (2006) showed that the effects of frame dragging on spectral lines are small for the 300--700~Hz spin frequencies observed from X-ray bursters and millisecond pulsars. However, stars spinning above 300~Hz become oblate in shape and acquire a non-zero quadrupole moment (e.g., Cook et al.\ 1994; see also Laarakkers \& Poisson~1999; Morsink et al.\ 2007, Pappas \& Apostolatos 2012). For such stars, the Schwarzschild+Doppler method is no longer accurate (Chang et al.\ 2006). Instead, a different metric is needed that takes into account higher order effects on the neutron star spacetime. 

For moderately spinning stars ($<$ 800~Hz), the external spacetime can be approximated analytically to second order in spin frequency by the Hartle-Thorne metric, which allows for an arbitrary quadrupole moment (Hartle \& Thorne 1968). The spacetimes around more rapidly rotating neutron stars can only be modeled with numerical metrics (see, e.g., Cook et al.\ 1994; Stergioulas \& Friedman 1995).

In this paper, we calculate the profiles of atomic features from neutron-star surfaces in the Hartle-Thorne approximation (Baub\"ock et al.\ 2012). This approach allows us to describe the external spacetimes of rotating neutron stars using only a few  of their macroscopic properties, such as their masses, radii, angular momenta, and quadrupole moments, without a detailed knowledge of the equation of state; naturally, these properties can be calculated for any given equation of state. We find that taking into account the quadrupole moment of the neutron-star spacetime leads to atomic features that are qualitatively different than those calculated in the Schwarzschild+Doppler approximation. Indeed, atomic features with very narrow cores may originate on the surfaces of even moderately spinning neutron stars when viewed at inclinations at which pure Doppler effects would predict significantly broader features. 

\section{Ray Tracing}

We describe the external spacetime around a neutron star by a variant of the Hartle-Thorne metric (Hartle \& Thorne 1968) constructed by Glampedakis \& Babak (2006). The Hartle-Thorne metric is a multipole approximation of the spacetime around a neutron star truncated at second order in spin frequency. The metric includes terms up to the quadrupole moment, while all higher-order moments are set to zero. The Glampedakis \& Babak metric we employ for this study is an expansion around the Kerr metric that allows for the quadrupole moment to be set to an appropriately chosen value. The higher order moments are equal to their Kerr values. The Hartle-Thorne and Glampedakis-Babak metrics are formally equivalent to second order in spin. 

We write the metric in Boyer-Lindquist coordinates as a deviation from the Kerr metric:
\begin{equation}
g_{\mu\nu}=g_{\mu\nu}^{\rm K}+\eta a^2 h_{\mu\nu},\;
\label{eq:gmunu}
\end{equation}
where $g_{\mu \nu}^{\rm K}$ is characterized by the line element

\begin{eqnarray}
ds^2&=&-\left(1-\frac{2Mr}{\Sigma}\right) dt^2 - \left(\frac{4Mar\sin^2\theta}{\Sigma}\right) dt d\phi\nonumber\\
&&+ \left(\frac{\Sigma}{\Delta}\right) dr^2 + \Sigma d\theta^2\nonumber\\
&&+ \left(r^2 + a^2 + \frac{2Ma^2r\sin^2\theta}{\Sigma}\right)\sin^2\theta d\phi^2,
\label{eq:dsK}
\end{eqnarray}
with
\begin{equation}
\Sigma\equiv r^2+a^2\cos^2~\theta\;
\label{eq:sigma}
\end{equation}
and
\begin{equation}
\Delta\equiv r^2-2Mr+a^2.
\label{eq:delta}
\end{equation}
Here $M$ is the mass of the neutron star and $a$ its specific angular momentum per unit mass.
 
The parameter $\eta$ measures the deviation of the quadrupole moment, $q$, of the neutron star from the Kerr value:
\begin{equation}
q = - a^2 (1 + \eta).
\label{eq:q}
\end{equation}
In the Kerr metric, $\eta = 0$ and Equation~(\ref{eq:q}) reduces to $q = -a^2$. The correction to the Kerr metric is given by \begin{eqnarray}
h^{tt}&=&(1-2M/r)^{-1}\left[\left(1-3\cos^2\theta\right)
\mathcal{F}_1(r)\right],\nonumber\\
h^{rr}&=&(1-2M/r)\left[\left(1-3\cos^2\theta\right)\mathcal{F}_1(r)\right],
\nonumber\\
h^{\theta\theta}&=&-\frac{1}{r^2}\left[\left(1-3\cos^2\theta\right)
\mathcal{F}_2(r)\right],\nonumber\\
h^{\phi\phi}&=&-\frac{1}{r^2\sin^2\theta}\left[\left(1-3\cos^2\theta\right)
  \mathcal{F}_2(r)\right],\nonumber\\
h^{t\phi}&=&0\;,
\end{eqnarray}
where the functions $\mathcal{F}_{1,2}$ are given explicitly in Appendix~A of Glampedakis \& Babak (2006). 

Laarakkers \& Poisson (1999) calculated the value of $\eta$ numerically for several equations of state and spin frequencies. They found that Equation~(\ref{eq:q}) provides a good description of the quadrupole moment even for stars spinning near their breakup frequency and that, in general, the parameter $\eta$ varies between 1 and 6. For the purposes of the illustrative calculations presented in this paper, we have adopted $\eta = 3.3$, roughly corresponding to a 1.4 $M_\odot$ star with an FPS equation of state (Laarakkers \& Poisson 1999). We have also adopted a value of $a = 0.357$ for a spin frequency of 700~Hz. We adjusted this value linearly with spin frequency in the calculations for slower neutron stars.

More recently, Pappas \& Apostolatos (2012) showed that there were inaccuracies in the way that Laarakkers \& Poisson (1999) inferred the multipole moments from the asymptotic expansion of their metric. They calculated corrected values of the quadrupole moment and find new fit parameters $\eta$ for several equations of state. However, the differences in the value of $\eta$ they obtain are small (less than 3\%). We have therefore adopted the Laarakkers \& Poisson value for the purpose of this study in order to be compatible with our earlier results (Baub\"ock et al.\ 2012).

In principle, the higher-order moments included in our metric could affect the line profiles we calculate here. However, we expect this contribution to be small. The next multipole moment after the mass quadrupole is the S$_3$ current quadrupole moment, which is smaller than the quadrupole moment by a factor of $a$. This moment affects only the off-diagonal terms of the metric tensor and thus modifies the frame dragging around the neutron star. Since frame dragging has only a small effect on line profiles emitted at the neutron-star surface (Bhattacharyya et al.\ 2006), we expect the contribution of the $S_3$ moment to be negligible. The next moment above the $S_3$ current quadrupole is the mass octupole moment. This moment is smaller than the mass quadrupole by a factor of $a^2$, so we also expect its contribution to the line profile to be small.

Neutron stars spinning at moderate to fast spin frequencies also acquire an oblate shape (Cook et al.\ 1994). In our calculations, we allow for the surface of the neutron star to deviate from spherical in addition to adopting a non-zero quadrupole moment. Morsink et al. (2007) approximate the oblate shape of the stellar surface by expanding the radius $R(\theta)$ as a series of Legendre polynomials:
\begin{equation}
\frac{R(\theta)}{R_{\rm eq}}=1+\sum_{n=0}^N a_{2n}P_{2n}(\cos\theta)\;,
\label{eq:shape}
\end{equation}
where $P_{2n}$ is the Legendre polynomial of order $2n$ and the coefficients $a_{2n}$ can be represented by simple polynomial functions of the compactness of the neutron star and of its spin frequency. (The exact forms of the coefficients $a_{2n}$ can be found in Morsink et al.\ 2007.) They find that even the shapes of the most rapidly rotating stars can be accurately described by truncating the series at $N=2$. Since the Hartle-Thorne metric is formally of second order, we terminate the series in Equation~(\ref{eq:shape}) at $N=1$. 

In order to model the appearance of a neutron star to an observer at infinity, we use the ray-tracing code detailed in Baub\"ock et al.\ (2012). In this code, an image plane is located at a large distance from the neutron star where the spacetime is asymptotically flat. Photon trajectories originate on the image plane and are traced backward through the curved spacetime using a fourth order Runge-Kutta algorithm. If the photons intersect the neutron star surface, the algorithm returns the $\theta$ and $\phi$ coordinates on the stellar surface, the energy of the photon, and its angle with respect to the surface normal. Photons that pass beyond the star by 10\% of the distance to the observer are halted and discarded. For the purposes of this paper, we have assumed that the emissivity is constant over the surface of the neutron star and independent of the beaming angle of the photons. 

The ray-tracing code produces a map on the image plane of the ratio of the energy of the photon at its place of origin to its energy at infinity. In order to calculate the emission line profile for a distant observer, we use a Monte-Carlo algorithm to sample points across the image plane. At each point, we find the redshifted energy by interpolating the four nearest grid points of the ray-tracing simulation. Points falling in regions corresponding to photon paths that miss the neutron star are discarded.

Additional effects at the neutron star surface may broaden the observed spectral lines beyond the purely kinematic and gravitational effects considered here. Mechanisms such as pressure broadening, thermal Doppler broadening, and Zeeman splitting are all expected to contribute to the total width of observed lines. However, for moderately spinning stars such as those considered in this paper, the combined effect of these mechanisms is expected to be small compared to the rotational broadening (\"Ozel 2013). Therefore, we initially neglect such broadening mechanisms and only consider the effects of finite intrinsic line widths at the end of Section 3.

\section{Line Profiles}
\begin{figure}[tbp]
\psfig{file=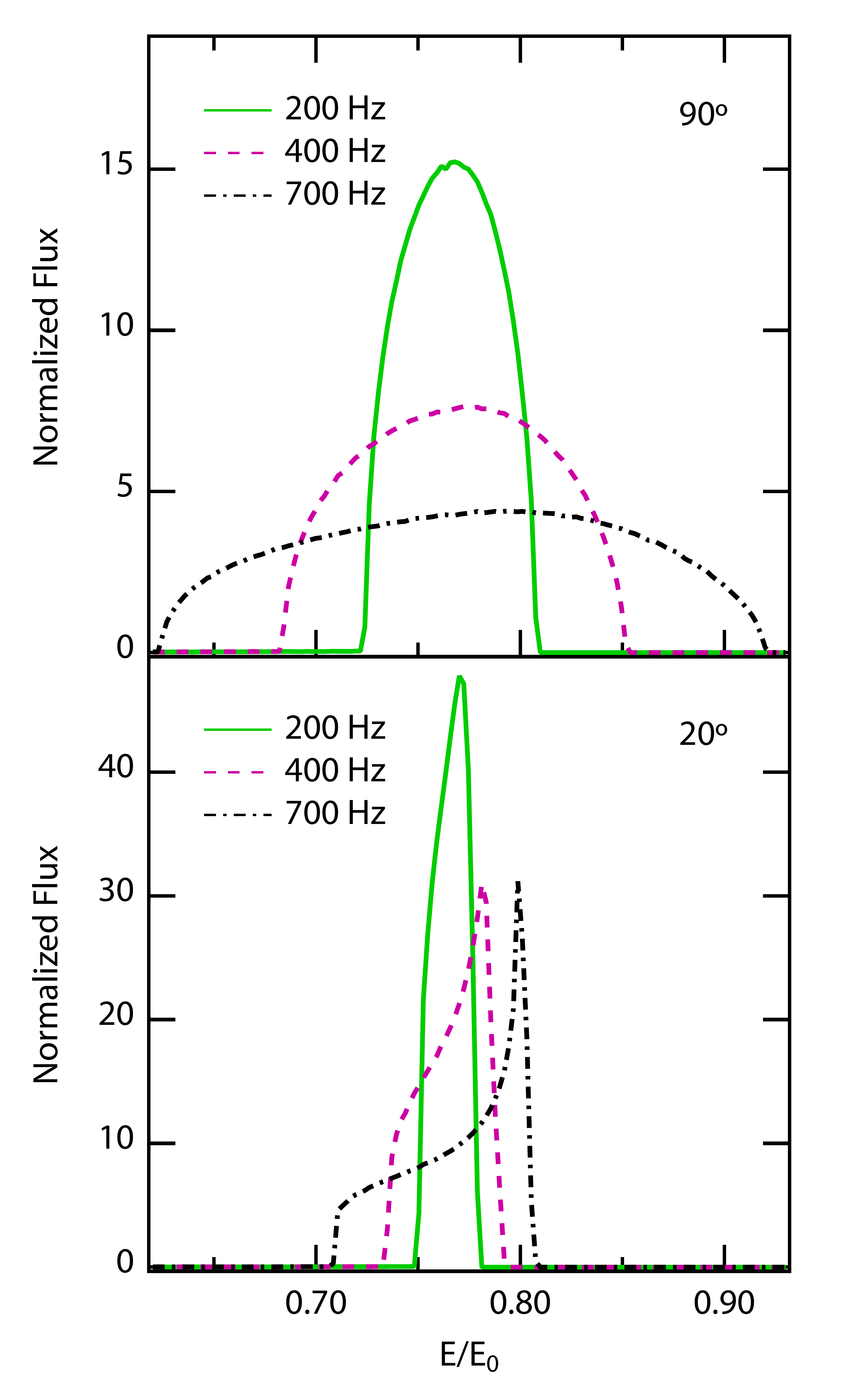,width=3.5in}
\caption{Line profiles for neutron stars spinning at three different rotational frequencies. In each case, the star has a radius of 10~km, a mass of 1.4~$M_\odot$, and an inclination of (top) 90$^\circ$ and (bottom) 20$^\circ$ to the observer. As the spin frequency increases, the lines become broader. At low observer inclinations, the lines also develop narrow cores.}
\label{fig:profiles}
\end{figure}

\begin{figure}[tbp]
\psfig{file=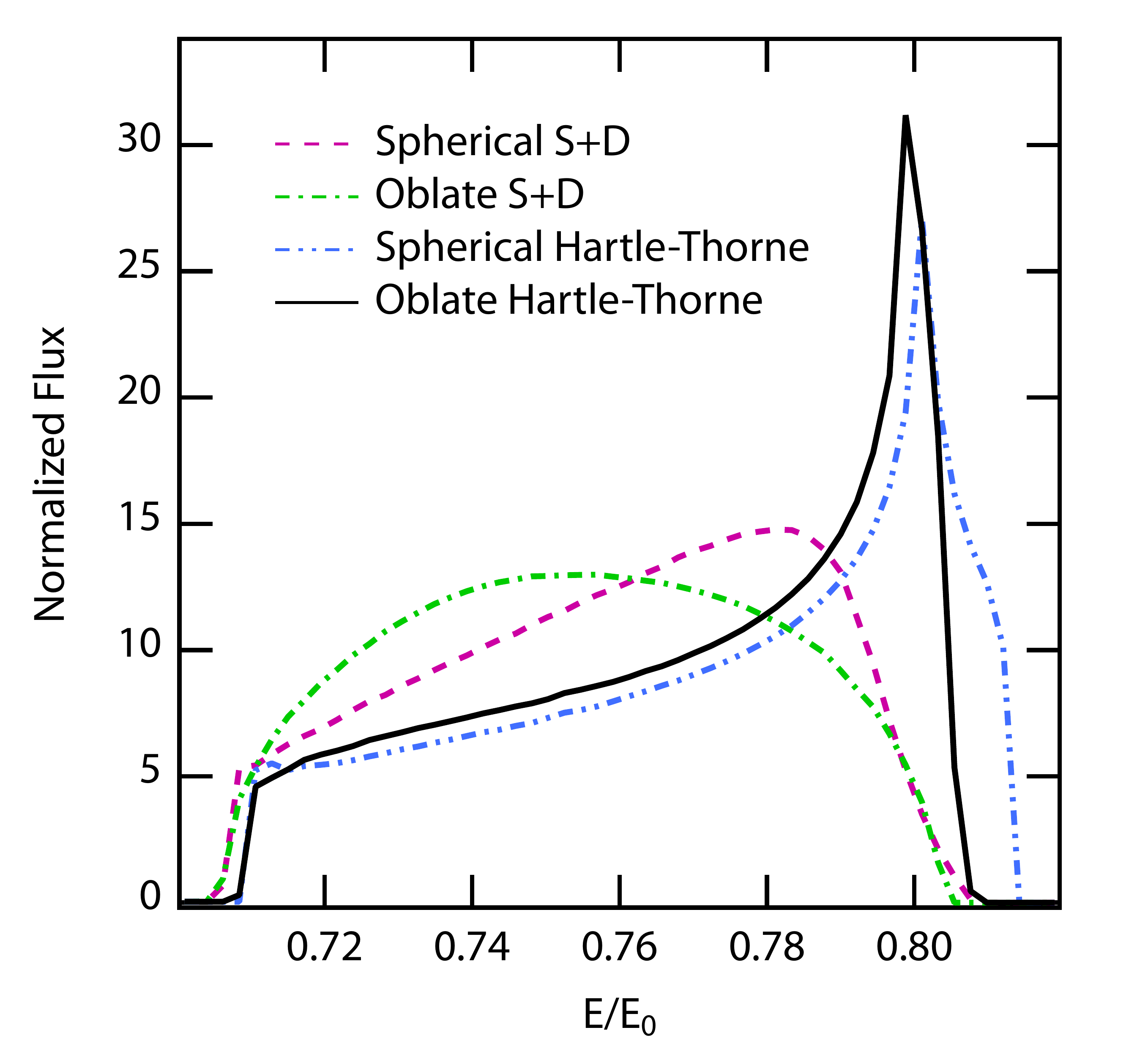,width=3.5in}
\caption{Line profiles computed by taking different properties of spinning neutron stars into account. In each case, the star is rotating at 700~Hz. As in Figure~\ref{fig:profiles}, the stars have a radius of 10~km, a mass of 1.4~$M_\odot$, and an inclination of 20$^\circ$. The four lines show the results of calculations in four different approximations: the Schwarzschild+Doppler approximation for a spherical and an oblate star, and the Hartle-Thorne approximation for a spherical and an oblate star. The last case is identical to the 700~Hz profile shown in Figure~\ref{fig:profiles}.}
\label{fig:q_o_profiles}
\end{figure}

\begin{figure}[tbp]
\psfig{file=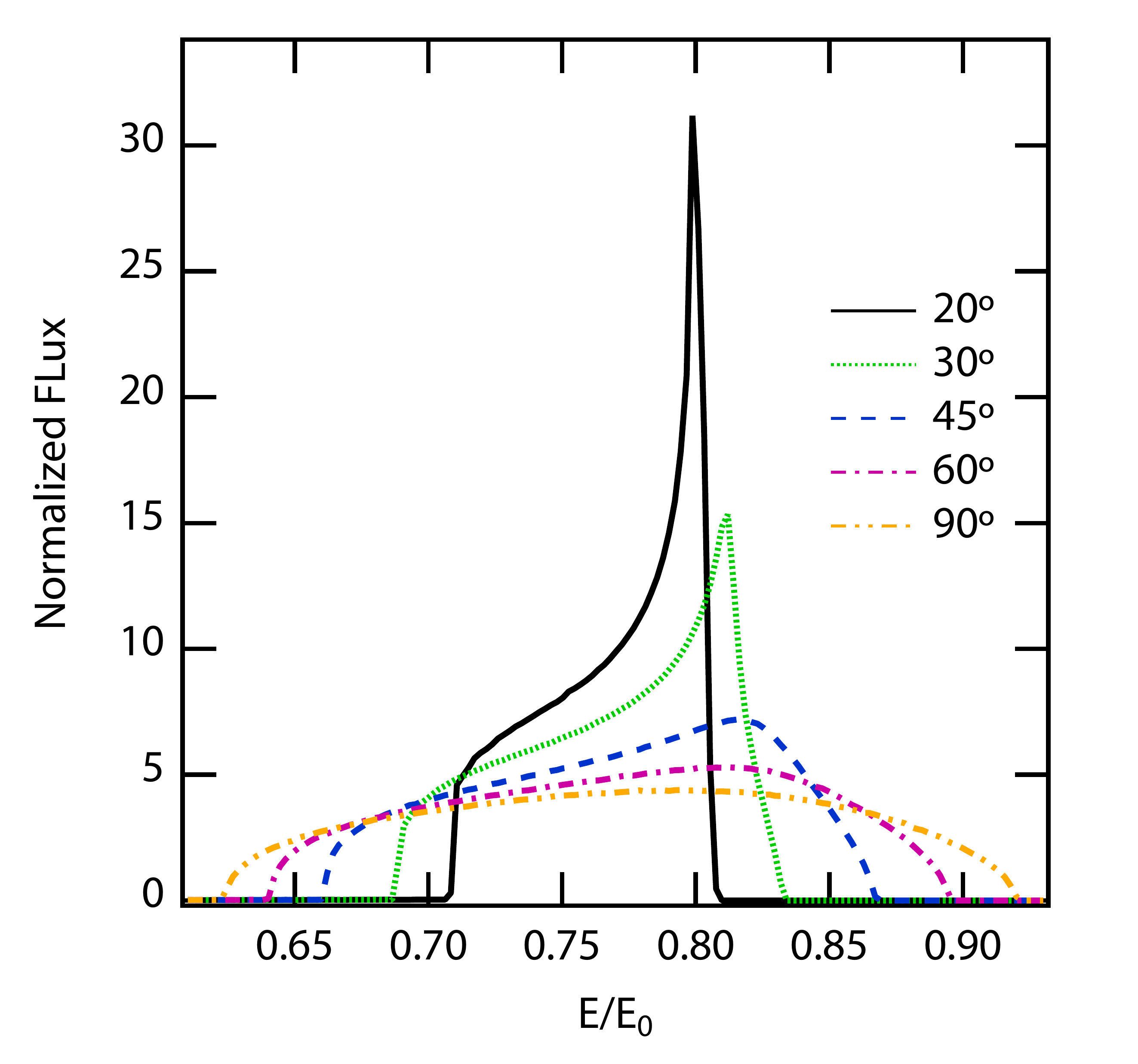,width=3.5in}
\caption{Line profiles for five different inclinations to the observer's line of sight. The neutron star has a radius of 10~km, a mass of 1.4~$M_\odot$, and a rotational frequency of 700~Hz. At high inclinations, the profile displays only Doppler broadening. At lower inclinations, however, a narrow peak appears.}
\label{fig:inc_profiles}
\end{figure}

\begin{figure}[tbp]
\psfig{file=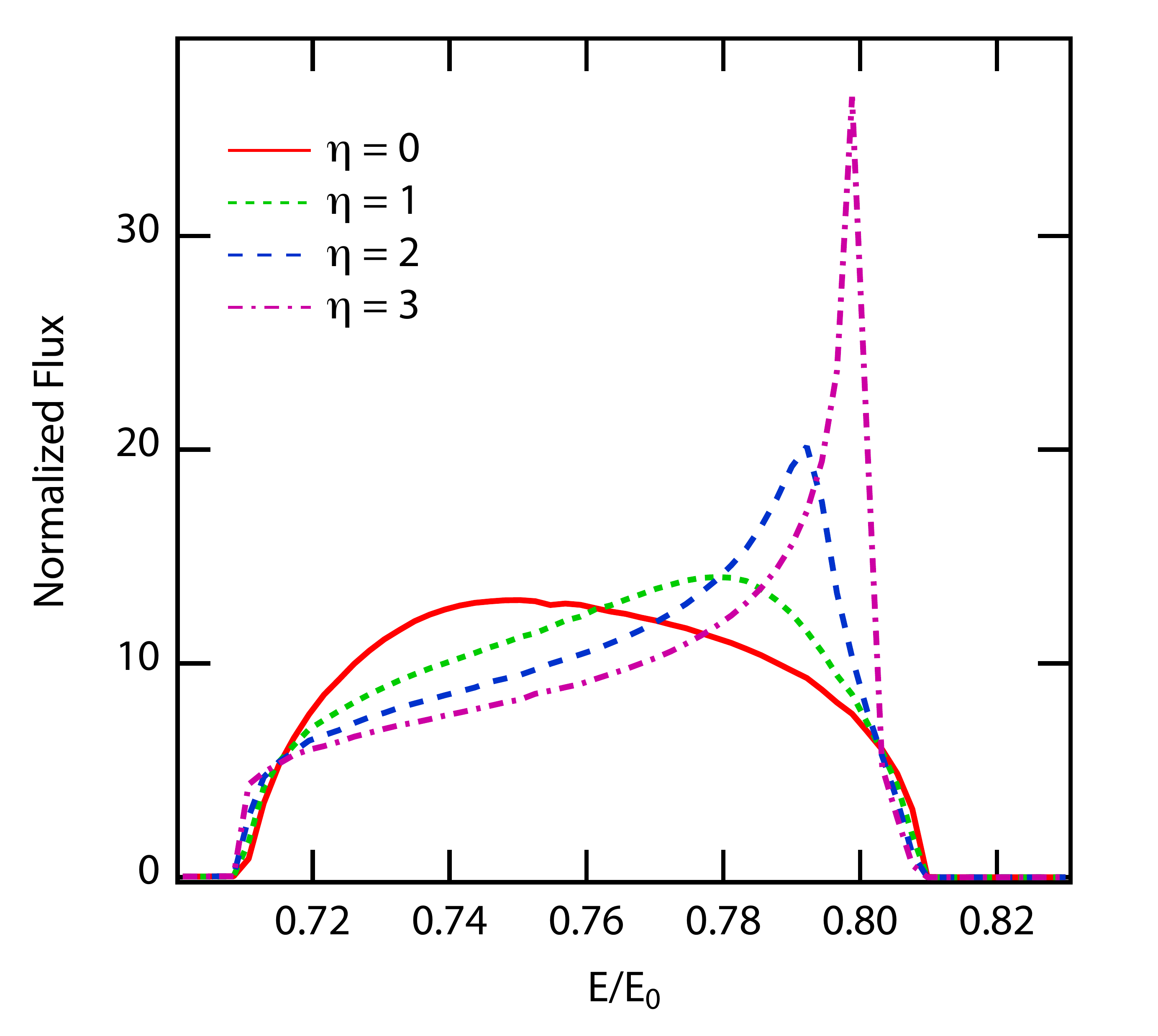,width=3.5in}
\caption{Line profiles for different values of the parameter $\eta$ which measures the quadrupole moment of the neutron star. As in previous figures, the neutron star has a 10~km radius, a mass of 1.4~$M_\odot$, and a rotational frequency of 700~Hz. The inclination to the observer's line of sight is 20$^\circ$. As the quadrupole moment (parametrized by $\eta$) increases, the line develops a narrow peak in the blueshifted portion of the spectrum.}
\label{fig:eta_profiles}
\end{figure}

\begin{figure*}[tbp]
\psfig{file=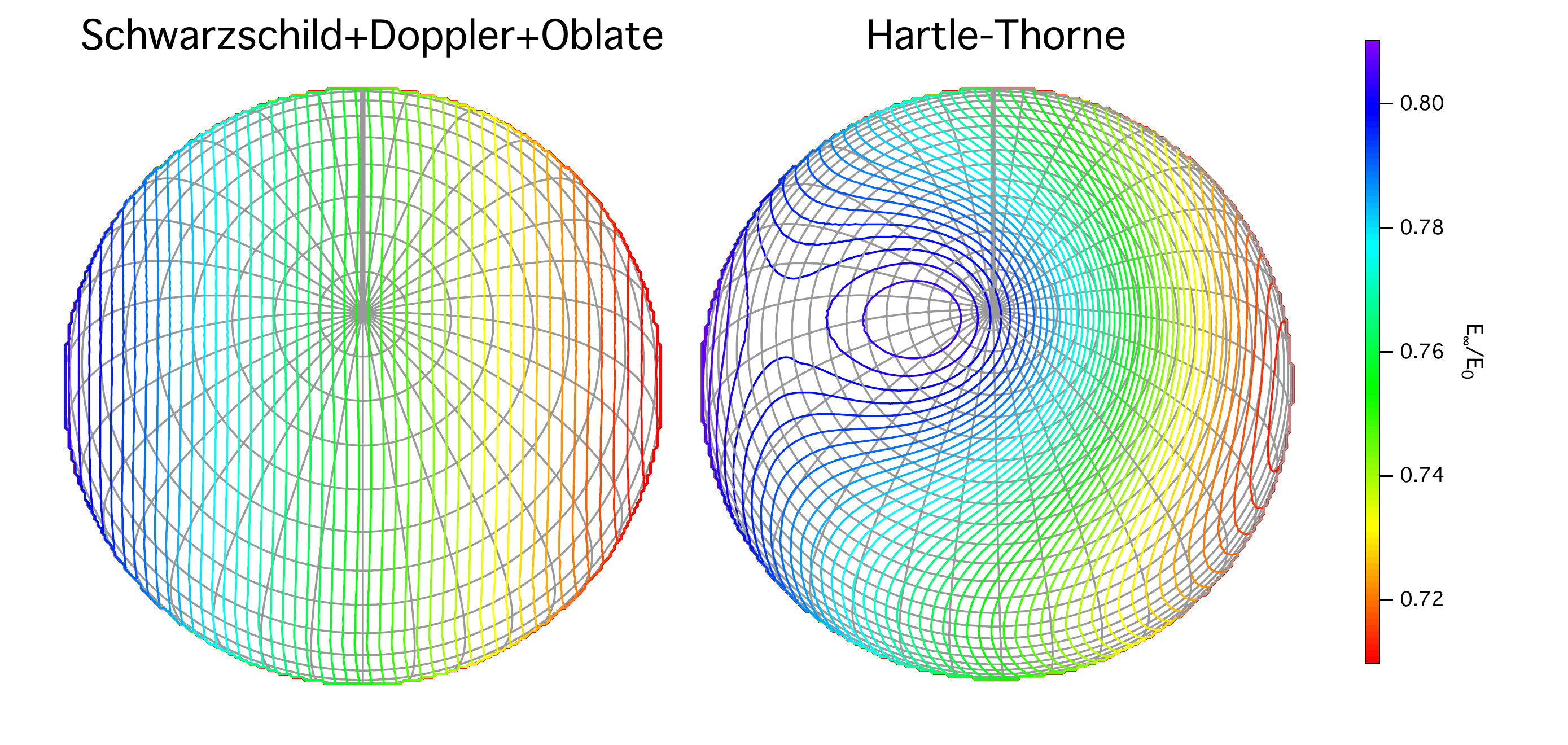,width=7in}
\caption{Contours of constant redshifted energy at the neutron star surface. The left panel shows the redshift for an oblate neutron star in the Schwarzschild+Doppler approximation, while the right panel includes our fiducial value for the neutron star's quadrupole moment as calculated by Laarakkers \& Poisson (1999). In both cases, the star has a mass of 1.4 $M_\odot$, a radius of 10~km, an inclination of $20^\circ$ to the observer's line of sight, and is rotating at 700~Hz. The star is rotating counterclockwise when viewed from above. In the right panel, the quadrupole moment of the neutron star causes a local extremum in the redshifted energy near rotational pole on the blueshifted side of the star. This maximum is responsible for the narrow core of the atomic features from moderately spinning neutron stars.}
\label{fig:E_Contour_Q_Effect}
\end{figure*}

Figure~\ref{fig:profiles} shows the profiles of atomic features emitted from the surface of a neutron star with a radius of 10~km and a mass of 1.4 $M_\odot$, for different spin frequencies and observer inclinations. For each simulated profile, we normalize the line so that the bolometric luminosity is equal to unity.

In the absence of rotation, the only effect the geometry of the star has on the line profile is to shift its observed energy by the gravitational redshift at the stellar surface. In the case of a 1.4 $M_\odot$ star with a radius of 10~km, the energy of the line center is redshifted by a factor of 0.77. 

The first order effect of rotation on the line profile is the broadening associated with the Doppler shift between the approaching and receding edges of the star. At low spin frequencies, the width of the line is determined almost entirely by this effect. A slight asymmetry appears in the profile that is due to the second order effect of relativistic beaming---the photons emitted on the blueshifted portion of the star are beamed toward the observer, shifting the peak of the observed spectrum to the right as shown in Figure~\ref{fig:profiles}. The profiles corresponding to high observer inclinations closely match the results of, e.g., \"Ozel \& Psaltis (2003), in which only first-order Doppler shift corrections were considered. 

At low inclinations and more moderate spins, we find a significant deviation from the line profile predicted by the Schwarzschild + Doppler broadening. In that simplified approach, the width of the line is determined by the line-of-sight velocity of the neutron star surface and scales with the sine of the inclination to the observer. We find, however, that neutron stars with relatively rapid rotations (such that their quadrupole moments are large) and low inclinations (such that the Doppler effects are small) generate line profiles with remarkably narrow cores that are strongly peaked toward the blue end of the spectrum. These peaks are evident in the higher frequency simulations shown in the bottom panel of Figure~\ref{fig:profiles}.

The deviation at low inclinations from the Schwarzschild+Doppler predictions is due primarily to the quadrupole moment of the neutron star, with a small additional effect from the oblate shape of the surface. To disentangle the relative roles of the Doppler shift, the quadrupole moment, and the oblateness, we simulated two different scenarios: one in which the profile is calculated in the Schwarzschild+Doppler approximation and the other in which the full Hartle-Thorne metric is used. In each case, we performed the calculations both for a star with a spherical surface and for one in which the surface becomes oblate. The results are shown in Figure~\ref{fig:q_o_profiles}. The profile labeled ``Spherical S+D'' shows only the expected Doppler broadening and relativistic beaming. When we allow for the shape of the star to become oblate, the profile shifts toward the red. This shift is caused by the increased gravitational redshift in the region near the pole. The effect of the oblate shape of the stellar surface is overpowered, however, by the contribution of a nonzero quadrupole moment. As shown in the two Hartle-Thorne scenarios illustrated in Figure~\ref{fig:q_o_profiles}, the quadrupole causes a narrow peak in the line profile for both the spherical and the oblate stellar surfaces.

This narrow peak vanishes at higher inclinations, as shown in Figure~\ref{fig:inc_profiles}. The peak is very strong at inclinations below $30^\circ$, but the effect of the quadrupole decreases rapidly as inclination increases because of the increasing relative importance of the Doppler broadening. When viewed perpendicular to the axis of rotation, the peak vanishes and the profile can be described purely by Doppler broadening. Moreover, the narrow core has a strong dependence on the magnitude of the quadrupole moment of the neutron star. Figure~\ref{fig:eta_profiles} shows line profiles for four different values of $\eta$, ranging from 0 (the Kerr case) to 3. As the quadrupole moment grows, the peak in the line profile becomes more pronounced. At the quadrupole moment predicted for this neutron star configuration, the narrow core dominates the line width.

The appearance of a narrow peak in our simulated spectra arises from the fact that the addition of a quadrupole moment causes a significant portion of the stellar surface to have roughly the same effective redshift. This effect is a combination of the varying Doppler shift across the surface and of the quadrupole component of the gravitational redshift. We show in Figure~\ref{fig:E_Contour_Q_Effect} contours of constant redshifted energy on the stellar surface for two stars: the right panel depicts a neutron star with our fiducial value of the quadrupole moment, while the left panel shows the same calculation performed in the Schwarzschild+Doppler metric for an oblate neutron star. On the star with zero quadrupole moment, the energy decreases monotonically toward the right (the receding side) due to the changing Doppler shift. In the other case, there is an additional effect due to the quadrupole moment that is symmetric about the pole of rotation. The result is a local maximum of energy that is offset slightly from the pole. The energy of this local maximum corresponds to the energy of the peak in Figure~\ref{fig:profiles}.

\begin{figure}[tb]
\psfig{file=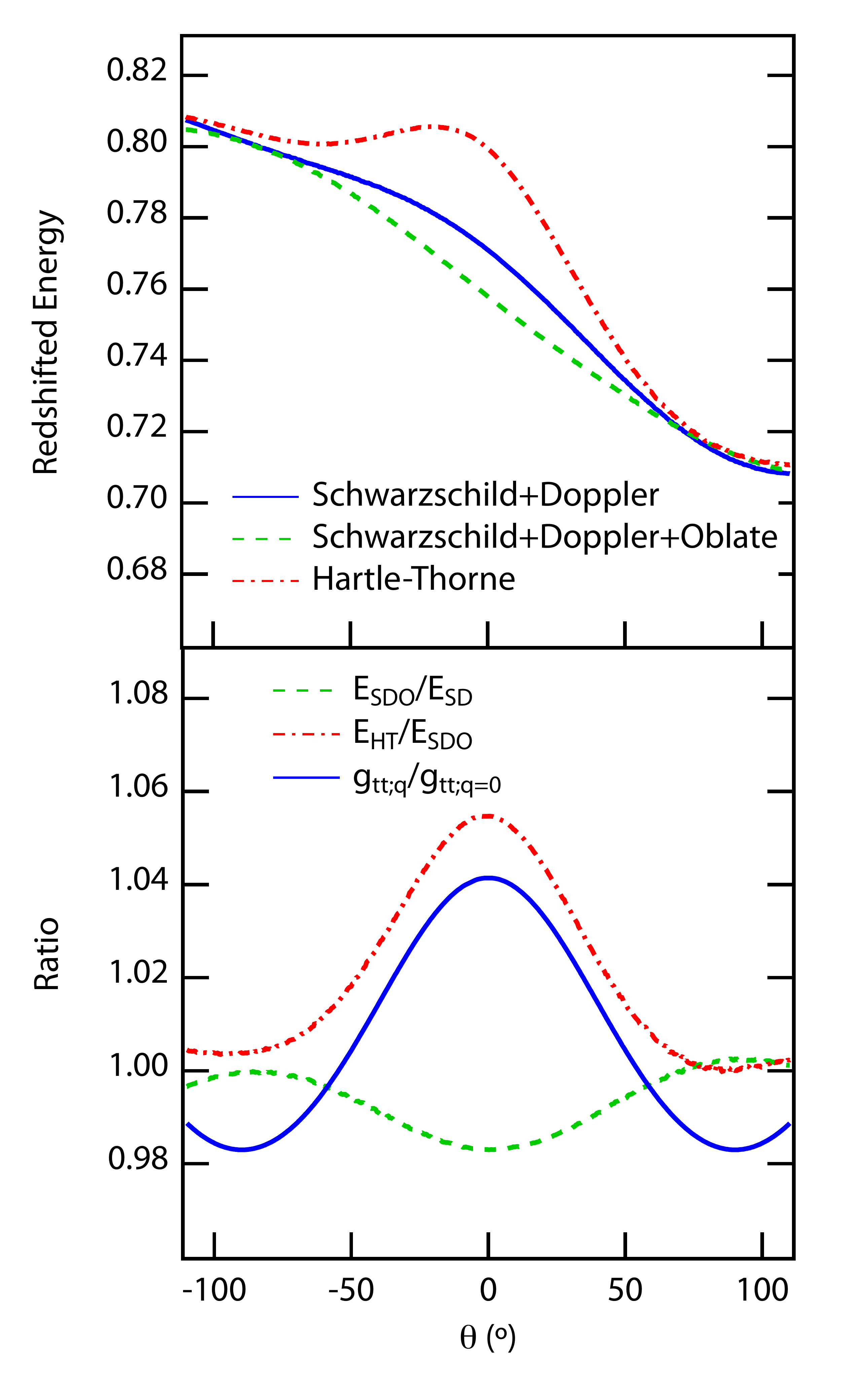,width=3.5in}
\caption{Top plot shows the redshifted energy of photons originating on the neutron star surface as a function of colatitude $\theta$ for a fixed value of the azimuth $\phi$ and for the neutron star parameters depicted in Figure~\ref{fig:E_Contour_Q_Effect}. The colatitude $\theta$ is zero at the rotational pole; it assumes negative values on the blueshifted side of the star and positive values on the redshifted side. The three lines correspond to three different configurations of the neutron star: one in which the star is spherical and has no quadrupole moment, one in which the star is oblate with no quadrupole moment, and one with both oblateness and a quadrupole moment. In the lower panel, the dashed line shows the ratio of redshifted energies on a spherical neutron star to those on an oblate one. The dash-dotted line shows the ratio of the redshifted energies of a neutron star with a quadrupole moment to those on a star without. The solid line shows the ratio $\lambda$ of the $g_{tt}$ components of two different spacetimes: one in which the quadrupole deviation is set to zero and another in which the quadrupole is set to its fiducial value.}
\label{fig:Q_Colat}
\end{figure}

In Figure~\ref{fig:Q_Colat}, we present another view of the combination of effects that give rise to these particular line profiles. The top panel shows the total redshifted energy as a function of colatitude for a constant value of the azimuth $\phi$ close to the local maximum for the configuration shown in the right panel of Figure~\ref{fig:E_Contour_Q_Effect}. For a spherical star with no quadrupole, the redshift changes monotonically across the surface due to the Doppler shift. An oblate star with no quadrupole moment is more redshifted in the vicinity of the pole, causing the shift in the line peak seen in Figure~\ref{fig:q_o_profiles}. In contrast, for an oblate star with the fiducial quadrupole moment, a local maximum appears that is offset from the pole. 

By taking various ratios of the curves shown in the upper panel of Figure \ref{fig:Q_Colat}, we can identify the effect of the oblateness and of the quadrupole moment. This is shown in the lower panel of Figure~\ref{fig:Q_Colat}. For comparison, we also calculate the ratio of the $g_{tt}$ terms of two metrics, one with our fiducial quadrupoles and the other with the quadrupole deviation set to zero:
\begin{equation}
\lambda = \frac{g_{tt, q}[\theta, R(\theta)]}{g_{tt, q=0}[\theta, R(\theta)]}
\label{gtt_ratio}
\end{equation} 
This ratio is shown as a dash-dotted curve in Figure~\ref{fig:Q_Colat}. The similarity between the energy ratios and the metric curve strongly suggests that the maximum in redshifted energy is caused by the quadrupole moment.

\begin{figure}[tb]
\psfig{file=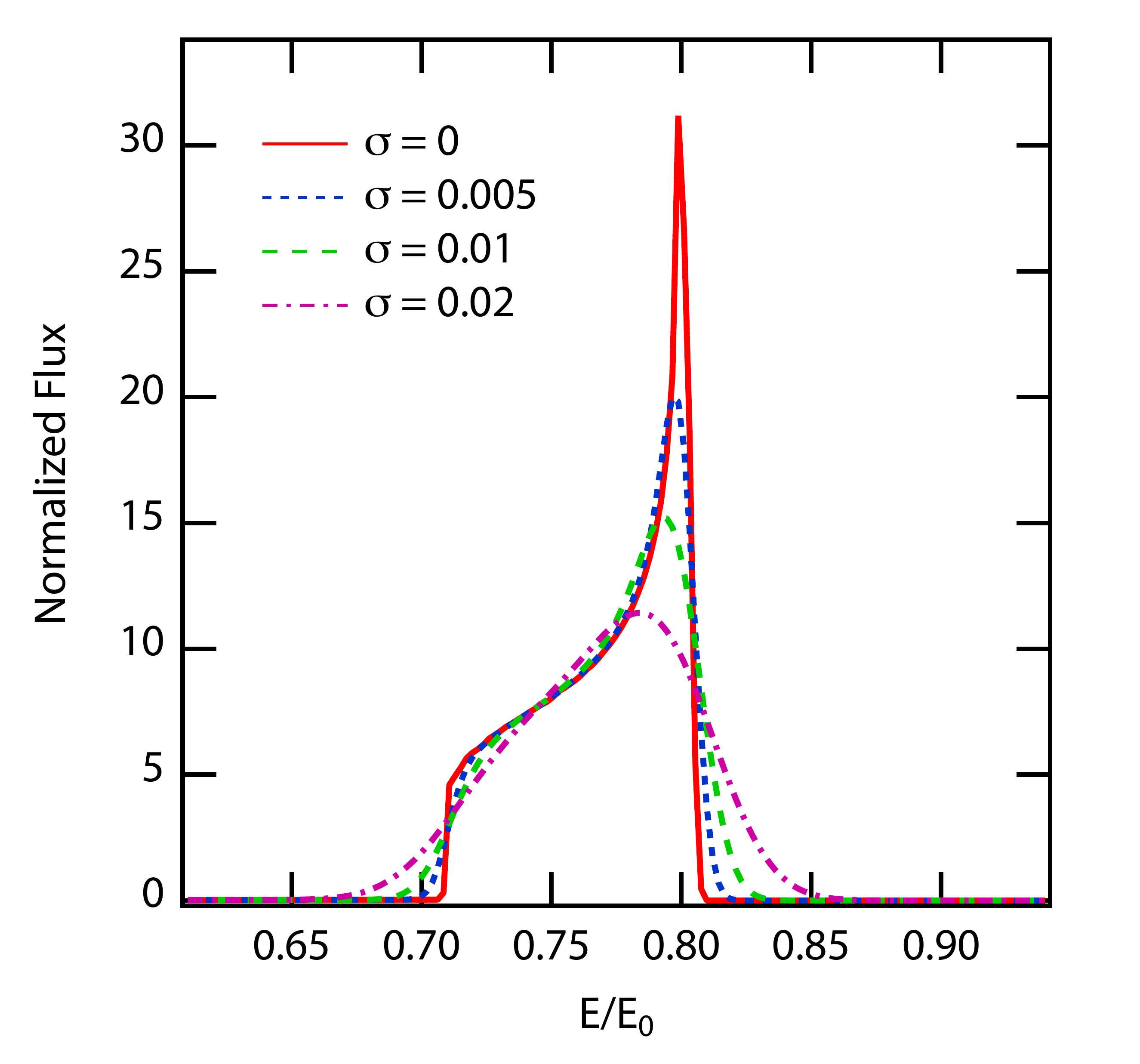,width=3.5in}
\caption{Line profiles for different intrinsic line widths. In each case, the result of the ray tracing for a star of 1.4 $M_\odot$ and a radius of 10~km spinning at 700~Hz is convolved with a Gaussian profile of width $\sigma$. The solid red profile shows a line with an infinitesimal intrinsic width and corresponds to the profile shown in Figures~\ref{fig:profiles}---\ref{fig:inc_profiles}. In order to illustrate the effect of intrinsic line broadening, we have included profiles corresponding to unrealistically broad lines. In practice, intrinsic line widths are expected to be at or below the $\sigma = 0.005$ level shown by the blue dotted curve.}
\label{fig:Gaussian_Broadening}
\end{figure}

Observed spectral lines from neutron star surfaces will be broadened by several effects in addition to the geometric effects considered above. In Figure~\ref{fig:Gaussian_Broadening}, we show the broadening of several Gaussian lines with different intrinsic widths. Models of neutron-star surface emission predict line widths on the order of $10^{-3}$ (\"Ozel~2013). As a result, the dotted line in Figure~\ref{fig:Gaussian_Broadening} shows a reasonable upper limit for the intrinsic line width expected from a neutron-star surface. For completeness, we also show profiles with significantly broader lines.

\section{Discussion}

\begin{figure}[tb]
\psfig{file=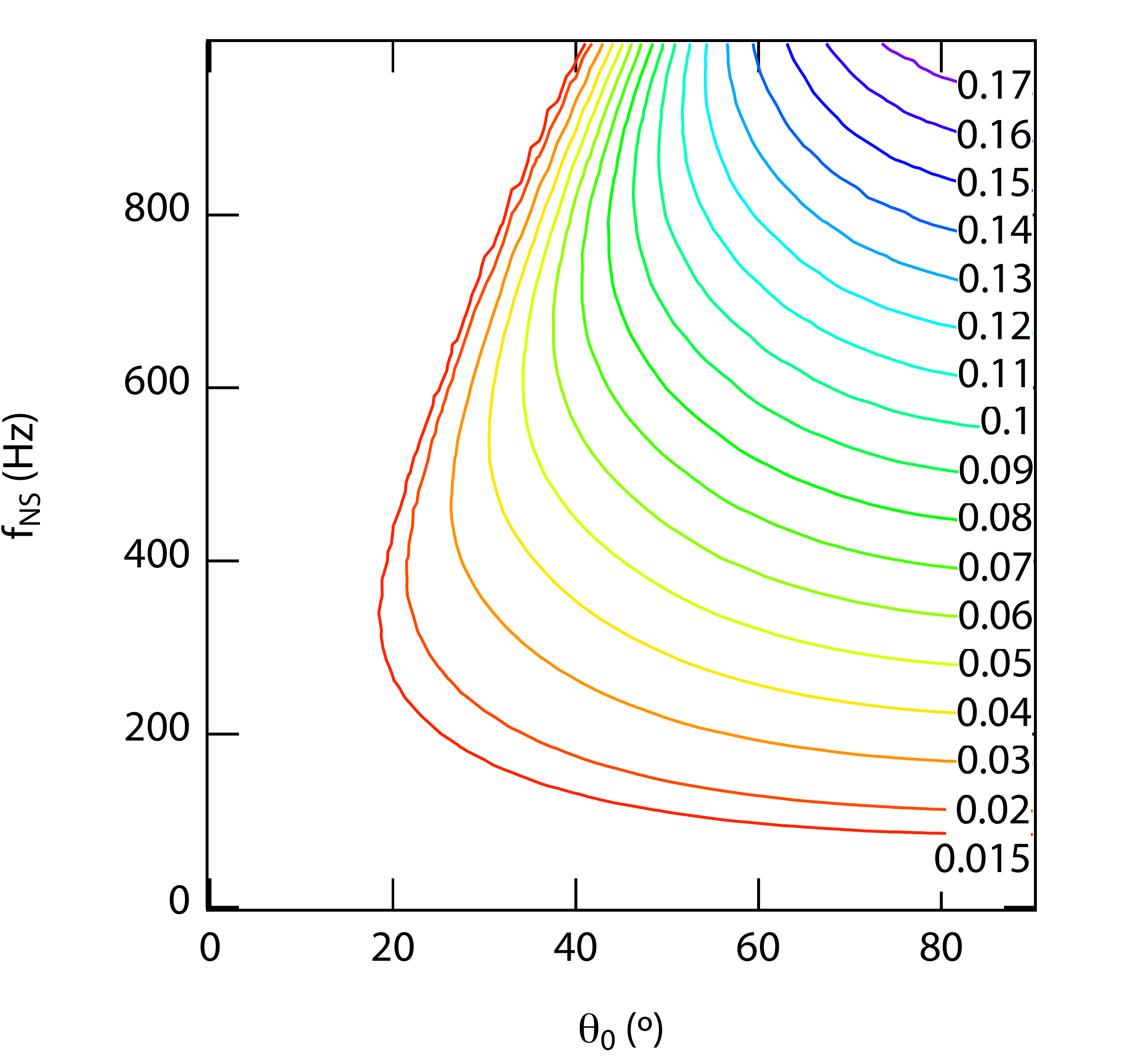,width = 3.5in}
\caption{Contour plot of the full width at half maximum of the line profile from a spinning neutron star of 1.4~$M_\odot$ and a radius of 10~km as a function of spin frequency and inclination angle.}
\label{fig:FWHM_Contour}
\end{figure} 

Second order effects of the rotation of neutron stars introduce significant corrections to the observed profiles of atomic features from their surfaces. The dependence of these profiles on  the inclination of the observer deviates strongly from the expected $\sin(\theta_0)$ behavior that is characteristic of Doppler effects. The principle cause of this deviation is the extra quadrupole mass moment induced by the rapid rotation of the star. The combination of the quadrupole moment and the Doppler shift due to the motion of the neutron star surface cause a strong and narrow feature to appear in the simulated spectra of emission lines. This narrow core is strongest at relatively low inclinations ($\theta_0~<~30^\circ$) and high spin frequencies ($f_{\rm NS}~>$~500~Hz).

Several previous authors have modeled the relativistic effects on spectral lines emitted from neutron star surfaces. \"Ozel \& Psaltis (2003) calculate line broadening in a Schwarzschild metric with Doppler shifts. Taking into account these first-order effects of rotation, they found uniformly broadened profiles, as expected from pure Doppler effects. At high inclination to the observer's line of sight, where quadrupole effects are small, our profiles are consistent with these calculations. 

Bhattacharyya et al.\ (2006) extended this analysis to include the effect of frame dragging. In their approximation, the authors assumed the neutron star to be spherical and the exterior spacetime to be described by the Kerr metric. Moreover, they assumed that line emission is confined to a band of latitudes on the surface of the neutron star. A neutron star in the Kerr metric has a small but non-zero quadrupole moment, so in principle narrow line cores could result in this approximation as well. However, because the quadrupole moment is small, the effect is only apparent at high spin frequencies, very low observer inclinations, and for bands of emission close to the rotational pole. The profiles in Bhattacharyya et al.\ (2006) are predominantly at high observer inclinations and for emission bands close to the rotational equator; therefore the effects of the Kerr quadrupole moment on their profiles is negligible.

Chang et al. (2006) calculated line profiles using numerical metrics for several equations of state. These numerical metrics incorporate appropriate quadrupole and higher order moments of the spacetime and should, therefore, reproduce the effects discussed here. However, the authors confined their analysis to high-inclination sources and relatively low spin frequencies, at which the contribution of the quadrupole moment to the line profile is small.

Figure~\ref{fig:FWHM_Contour} shows a contour plot of the simulated full width at half-maximum of emission lines for a range of spin frequencies and inclinations. At high inclinations, the interpretation of line widths is straightforward, with narrow lines corresponding to slow rotation and broad lines implying fast rotation. At lower inclinations, however, there is often a degeneracy. Narrow lines could be caused by slow rotation or by the higher quadrupole induced by rapid rotation. Clearly, line broadening can only be used as a reliable measure of neutron star radii if additional information is available on the inclination of the observer and the spin frequency.

The presence of a narrow core in atomic features also increases the range of inclinations at which these features are detectable. At high spin frequencies and relatively low observer inclinations, atomic features from realistic stars are narrower than those expected from purely Doppler effects (Chang et al. 2006). This may allow for such features to be discernible from continuum spectra and translates into a larger solid angle over which they are detectable

The models presented here may shed new light on observations of the source EXO 0748--676. {\em XMM} observations of this source showed evidence for narrow absorption lines (Cottam et al. 2002), which were later used to constrain the mass and radius of the neutron star (e.g., \"Ozel 2006). However, Lin et al. (2010) argued that the subsequent detection of the 552~Hz spin frequency and the large amplitude of burst oscillations were incompatible with the narrow observed width of the absorption lines and concluded that these lines did not originate at the neutron star surface.

In their calculations, Lin et al.\ (2010) use the Schwarzschild+Doppler approach to model the line spectra. As shown above, however, the inclusion of second-order effects leads to narrow profiles for much higher inclinations at a given spin frequency. Therefore, it appears plausible that the narrowness of the absorption lines is not incompatible with the high amplitudes of the observed burst oscillations and, therefore does not exclude the possibility that they were emitted from the surface. A detailed analysis of the EXO 0748--676 case will be presented in a future paper.

\acknowledgments

We thank the referee, Sharon Morsink, for many useful suggestions. We gratefully acknowledge support from NSF CAREER award AST-0746549, NSF grant AST-1108753, and Chandra Theory grant TM2-13002X for this work.

\end {document}